# IPv6 an IPv4 Threat reviews with Automatic Tunneling and Configuration Tunneling Considerations Transitional Model: -A Case Study for University of Mysore Network-

[1]Hanumanthappa.J.
[1]Dos in Computer Science,
University of Mysore,Manasagangothri,
Mysore,INDIA,
hanums_j@yahoo.com

Dr.Manjaiah.D.H.[2]
[2] CS at Dept.of Computer Science,
Mangalore University,Mangalagangothri,
Mangalore,INDIA.
ylm321@yahoo.co.in

***Abstract***

**The actual transition from IPv4 to IPv6 requires network administrators to become aware of the next generation protocol and the associated risk problems.Due to the scale and complexity of current internet architecture how to protect from the existing investment and reduce the negative influence to users and service providers during the transition from IPv4 to IPv6 is a very important future topic for the advanced version of an internet architecture.This paper summarizes and compares the IPv6 transition mechanism methods like Dual Stack,Tunneling issues like IPv6 Automatic tunneling and manually configured tunneling considerations, the IPv6 transition scenarios,IPv6 transition security problems,highlights IPv6 and IPv4 threat review with automatic tunneling and configuration tunneling considerations.In this paper we have proposed a transitional threat model for automatic tunneling and a configuration tunneling that could be followed by the University of Mysore(UoM),to estimate automatic tunneling and a manually configured tunneling threat review issues.Furthermore,there are different tunneling mechanisms such as:IPv6 over IPv4 GRE Tunnel,Tunnel broker,Automatic IPv4–Compatible Tunnel and Automatic 6-to-4 Tunnel and also outlines many of the common known threats against IPv6 and then it compares and contrast how these threats are similar ones,might affect an IPv6 network.**

## I. Introduction

In the last 20 years,the internet undertook a huge and unexpected explosion of growth [63].There was an effort to develop a protocol that can solve problems in the current Internet protocol which is in the Internet protocol version 4(IPv4).It was soon realized that the current internet protocol the IPv4,would be inadequate to handle the internet's continued growth.The internet Engineering task force(IETF) was started to develop a new protocol in 1990's and it was launched IPng in 1993 which is stand for Internet Protocol Next Generation.So a new generation of the Internet Protocol(IPv6)was developed [7],allowing for millions of more IP addresses.The person in charge of IPng area of the IETF recommended the idea of IPv6 in 1994 at Toronto IETF[1].But mainly due to the scarcity of unallocated IPv4 address the IPv4 protocol cannot satisfy all the requirements of the always expanding Internet because however its 32 bit address space being rapidly exhausted[2]alternative solutions are again needed[3].The long term solution is a transition to IPv6[5]which is designed to be an evolutionary step from IPv4 where the most transport and application–layer protocol need little or no modification to the work.The deployment of NAT[3]can alleviate this problem to some extend but it breaks end to end characteristic of the Internet,and it cannot resolve the problems like depletion(exhaustion) of IPv4 addresses.IPv6 protocol has 128-bit addresses instead 32 bit IPv4 addresses,however the migration from IPv4 to IPv6 is an instant is impossible because of the huge size of the Internet and of the great number of IPv4 users[16].Moreover, many organizations are becoming more and more dependent on the Internet for their daily work,and they therefore cannot tolerate downtime for the replacement of the IP protocol.IPv6 has some transition methods or techniques that permit end user to put into operation slowly but surely provides a high level of interoperation between both protocols IPv4 and IPv6.The IPv6 has some transition methods or techniques that permit end user to put into operation IPv6 slowly but surely and provides a high level of interoperation between both the protocols IPv4 and IPv6.The current IPv4 based Internet is so large and complex that the migration from IPv4 to IPv6 not as simple as the transition from NCP network to TCP/IP in 1983and also will take so many years to occur very smoothly[63].

can install it as a software which upgrade in most Internet machines,and it can work smoothly with the current IPv4 data.

This Journal paper is intended to provide a review of the most common threats in IPv6.This paper also can extend by investigating the additional threats through the deployment of IPv6 including those associated with the various transition mechanisms available with specific emphasis on Automatic tunneling and Configuration tunneling.Since there are no transitional networks(ISP's)IPv6 ready yet in INDIA,therefore,to get going deploying IPv6 we need transition techniques,since there is no complete world wide IPv6 network infrastructure.We should look at this stage as strategic vision, and we should look at the existing of IPv4 and IPv6 network infrastructure as necessary situation before complete migration to IPv6.The scarcity of information on the subject of IPv6 migration costs,merged with the reality that many organizations are not sold on the supposed benefits offered by the IPv6,is making the case for upgrading difficult to argue[2].It is quite obvious that changing from IPv4 to IPv6 is very costly,since many current network applications running on IPv4 .

This paper presents a comprehensive explanation about the current status of research on IPv6 Transition mechanisms,Tunneling types like Automatic Tunneling and Manually configured tunneling etc,Tunneling types threat reviews,IPv6 Security aspects,Threat review model and indicates the prospect of the future research.The paper is organized as follows: We briefly described the Theoretical considerations of IPv6 Transition issues in Section 2.We described IPv6 to IPv4 threat review in section 3.We discuss a brief overview of IPv6 Automatic Tunneling and Configuration tunneling mechanism considerations in section 4.We discussed prototype which explains the Automatic tunneling,Configuring tunneling review research in section 5.We describe our research approach recommendations on IPv6 tunneling threat types in section 6.In section 7 we will learn the current and future Innovative research challenges of IPv6 threat issues for researchers,finally we concluded the whole paper in section 8.

## II.Theoretical Consideration.

*A .Types of Transition Strategies in IPv6*

Tunneling is a strategy used when two computers using IPv6 want to communicate with each other and the packet must pass through a region that uses IPv4.To pass through this region,the packet must have an IPv4 address.So the IPv6 packet is encapsulated in an IPv4 packet when it enters

The key elements of these transition technologies are dual stack and configuration tunneling.The below figure-1 shows description of the different IPv6 tunneling scenarios and their configurations which are explained by using some of the available commands.The main important IPv6 transition techniques are Dual-Stack,Tunneling Techniques,and Header translation.

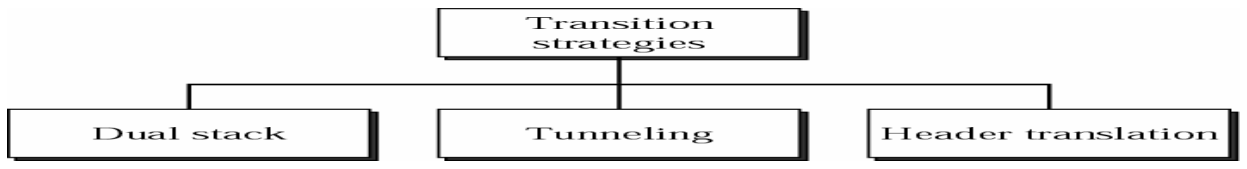

Fig .I.The IPv6 Transition Mechanisms.

*B. Introduction to Tunnels*

A tunnel is a bidirectional point –to-point link between two network endpoints.Data is carried through the tunnel using a process called encapsulation in which IPv6 packet is carried inside an IPv4 packet which makes IPv4 as a Data Link layer with respect to IPv6 packet transport. The term "tunneling" refers to a means to encapsulate one version of IP in another so the packets can be sent over a backbone that does not support the encapsulated IP version. For example,when two isolated IPv6 networks need to communicate over an IPv4 network,dual-stack routers at the network edges can be used to set up a tunnel which encapsulates the IPv6 packets within IPv4,allowing the IPv6 systems to communicate without having to upgrade the IPv4 network infrastructure that exists between the networks. This mechanism can be used when two nodes that use same protocol wants to communicate over a network that uses another network protocol.The tunneling process involves three steps:encapsulation,decapsulation,and tunnel management.It also requires two tunnel end-points,which in general case are dual-stack IPv4/IPv6 nodes,to handle the encapsulation and decapsulation.There will be performance issues associated with tunneling,both for the latency in en/de capsulation and the additional bandwidth used.Tunneling is one of the key deployment strategies for both service providers and enterprises during the period of IPv4 and IPv6 coexistence.Fig-2 Shows the deployment of IPv6 over IPv4 tunnels.

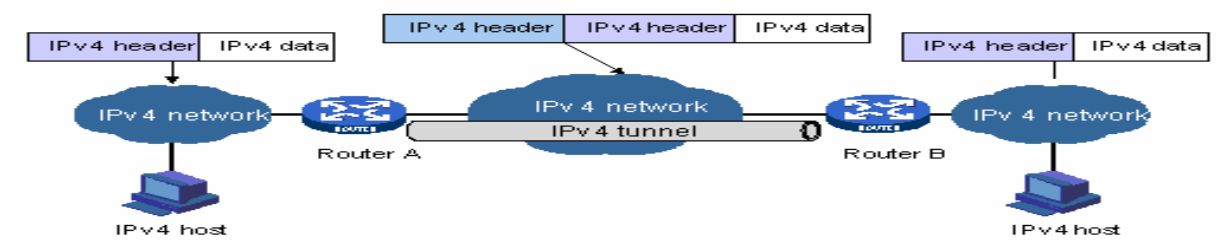

Fig.II.The Deploying an IPv6 over IPv4 Tunnels.

the region,and it leaves its capsule when it exists the region.It seems as if the packet IPv6 packet goes through a tunnel at one end and emerges at the other end [10].To make it clear that the IPv4 packet is carrying an IPv6 packet as data,the protocol value is set to 41[see Figure-2]Although IPv6 Dual Stack,IPv6 Tunneling,IPv6 Header Translation,are providing with us with transition solution but still it is not complete,there are still some other issues we should consider to get complete solution for transitioning.Tunneling techniques are broadly divided into two types,first one is an automatic tunneling and second one is configuration tunneling[Ref-Figure-4].The tunneling technique we can use the compatible addresses discussed as shown in the below Figure-3.A compatible address is an address of 96 bits of zero followed by 32 bits of IPv4 address.It is used when a computer using IPv6 wants to send a message to another computer using IPv6.However suppose the packet passes through a region where the networks are still using IPv4.The sender must use the IPv4-compatible address to facilitate the passage of the packet through the IPv4 region.For example the IPv4 address 2.13.17.14 becomes 0::020D:110E.The IPv4 is prepended with 96 zeros to create a 128–bit address (See figure-3)[10].

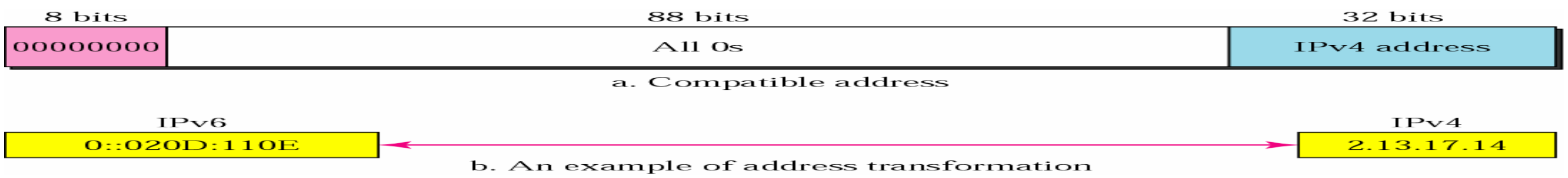

Fig.III.The IPv6 Compatible Address.

*C. Types of Tunneling Mechanisms.*

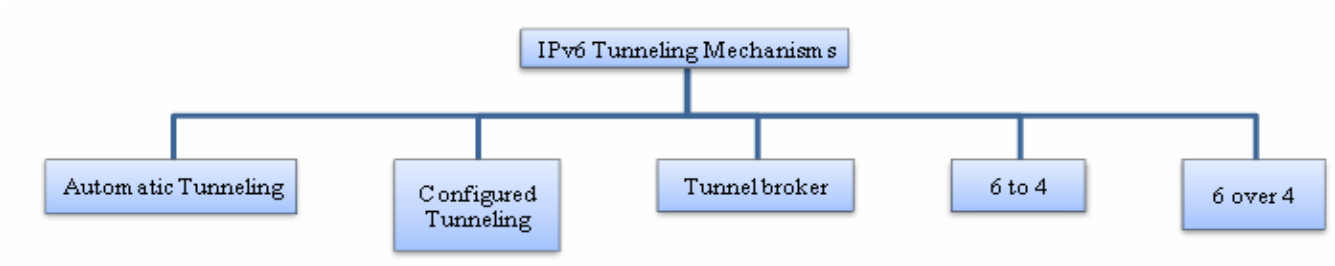

Fig.IV.The IPv6 Tunneling Mechanisms.

*D.An Automatic Tunneling*

If the receiving host when it uses a compatible IPv6 address,tunneling occurs automatically without any reconfiguration.In automatic tunneling,the sender sends the receiver an IPv6 packet using the IPv6 compatible address as the destination address.When the packet reaches boundary of the IPv4 network,the router encapsulates it in an IPv4 packet,which should have an IPv4 address.To get this address,the router extracts the IPv4 address embedded in the IPv6 address.The packet then travels the rest of its journey as an IPv4 packet.The destination host,which is using a dual stack,now receives an IPv4 packet.Recognizing its IPv4 address, it reads the header,and finds that the packet is carrying an IPv4 packet.It then passes the packet to the IPv6 software for processing.(SeeFigure-5)[10].

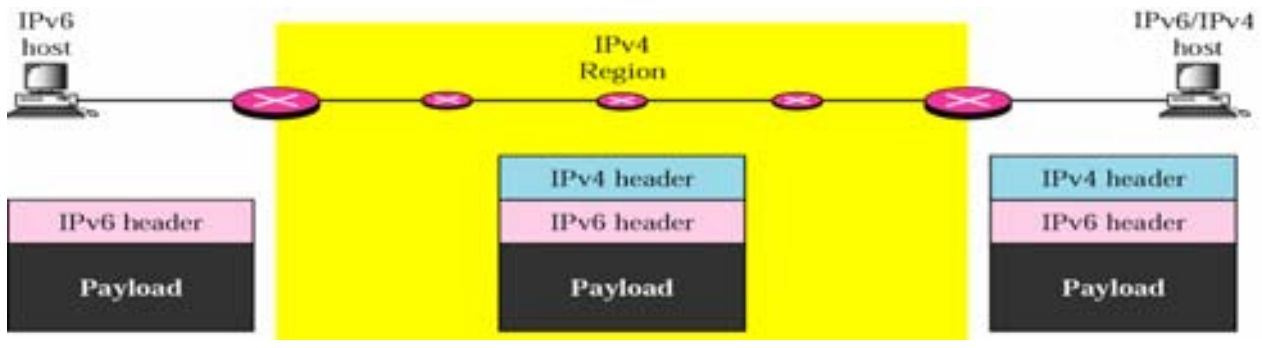

Fig .V.The IPv6 an Automatic Tunneling..

*E.Configured tunneling*

The tunnel point addresses are determined by the configuration information that is stored at the encapsulating end point hence the name configured tunneling.The other name for a configured tunneling is an explicit tunneling.If the receiving host does not support an IPv6 –compatible address,the sender receives no compatible IPv6 address from the DNS.In a configuration the sender sends the IPv6 packet with the receiver's no compatible IPv6 address, however the packet cannot pass through the IPv4 region without first being encapsulated in an IPv4 packet.The two routers at the boundary of the IPv4–region are configured to pass the packet encapsulated in an IPv4 packet.The router at one end sends the IPv4 packet with its own IPv4 address as the source address and the other router's address as the destination.The router receiver's the packet,decapsulates the IPv6 packet, and sends it to the destination host.The destination host then receives the packet in IPv6 format and processes it [10](SeeFigure-6).

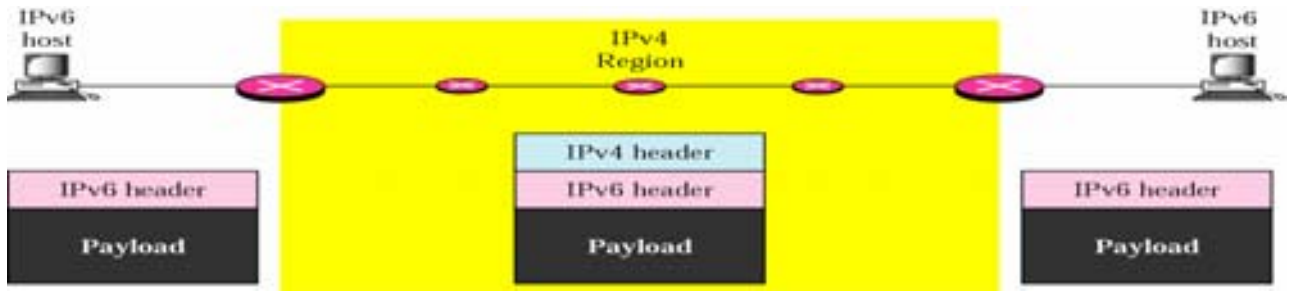

Fig .VI.The IPv6 Configuration Tunneling.

Configuration and Automatic tunnels can be defined to go between router-to-router,Host–to-Host,Host-to-Router,and Router–to-Host but are most likely to be used in a router–to-router configuration.

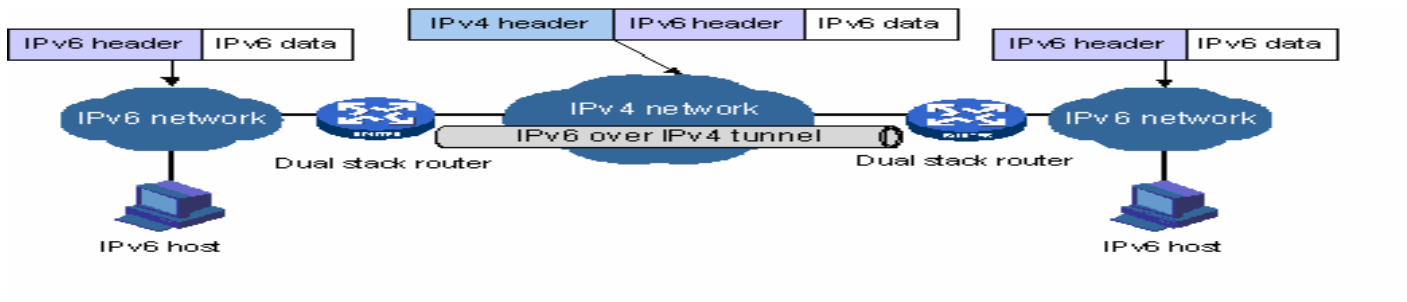

Fig.VII.The IPv6 over IPv4 GRE Tunneling.

The IPv6 over IPv4 GRE tunnel is a variety of tunneling mechanism in a TCP/IP protocol suite.This is also a type of GRE tunneling technique that is designed to provide the services necessary to implement standard point-to-point encapsulation scheme.GRE tunnels are the links between end points with a separate tunnel for each link as similar to IPv6 manually configuration tunnel.However each tunnel is not tied to a specific passenger or transport protocol but in this

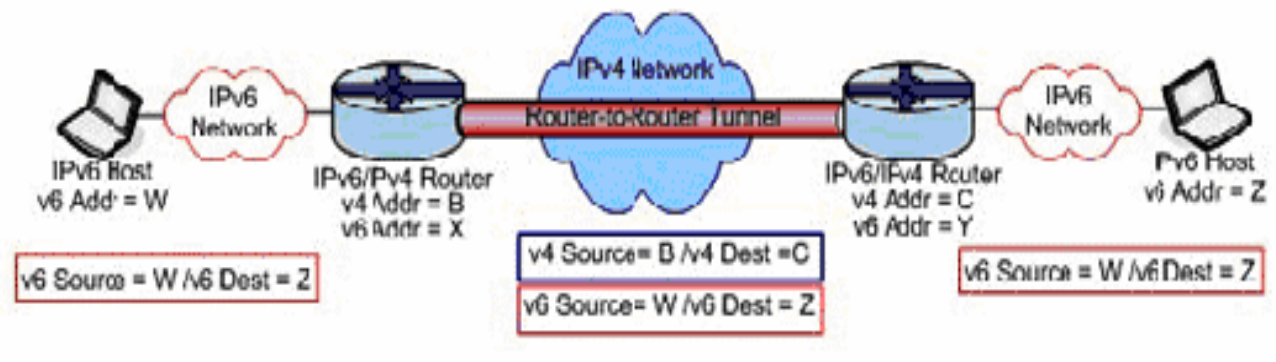

Fig.VIII. The Router –to-Router Tunneling Configuration.

3.Host-to-Host:IPv4/IPv6 hosts that are interconnected by an IPv4 infrastructure can tunnel IPv6 packets between themselves.

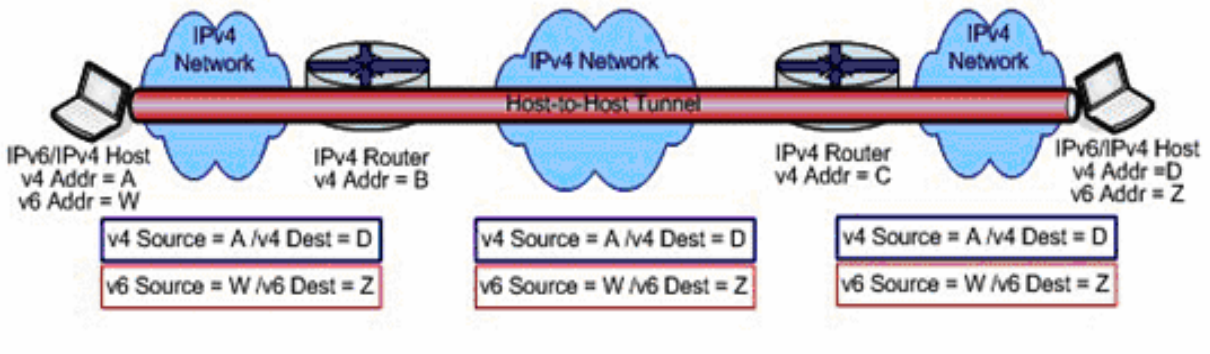

Fig.X.The Host–to-Host Tunneling Configuration.

## III.The IPv6 to IPv4 Threat review

### *A .Types of Threats in IPv6 Security*

### *F.IPv6 over IPv4 GRE Tunneling.*

situation they carry IPv6 as the passenger protocol over GRE as the carrier protocol.The Fig-7 Shows how to configure IPv6 over IPv4 GRE tunnel.

### *G.Configuration Tunneling Scenarios:*

1.Router –to-Router Tunneling Configuration: During the migration,the tunneling technique can be used in the following ways:

1.Router-to-router:IPv6/IPv4 routers interconnected by an IPv4 infrastructure can tunnel IPv6 packets between themselves.

2.Host–to-Router:IPv6/IPv4 hosts can tunnel IPv6 packets to an intermediary IPv6/IPv4 router that can be reached via an IPv4 Infrastructure.

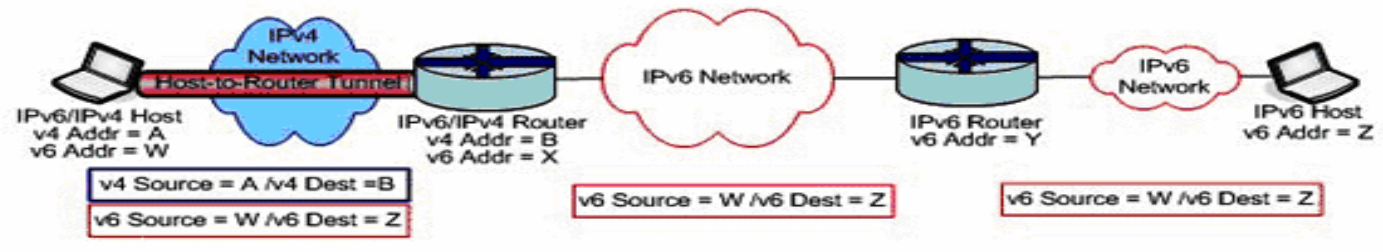

Fig.IX.The Host–to-Router Tunneling Configuration.

4.Router-to-Host:IPv6/IPv4 routers can use tunnels to reach an IPv4/IPv6 host via an IPv4 infrastructure.

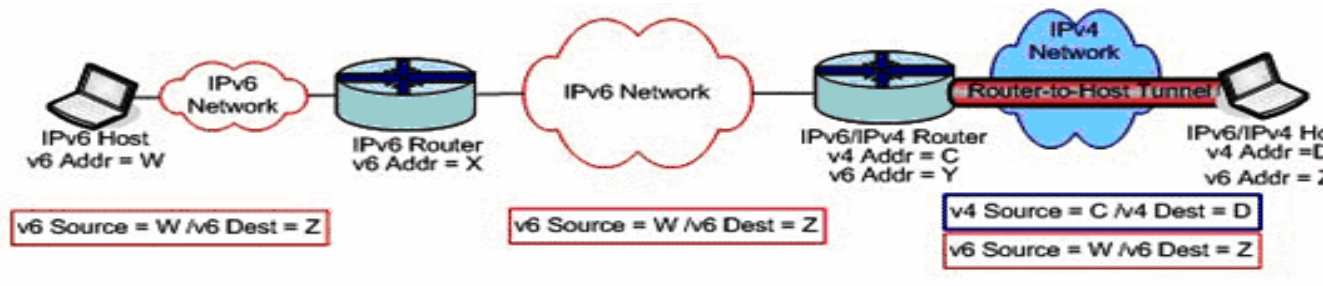

Fig.XI.The Router–to-Host Tunneling Configuration.

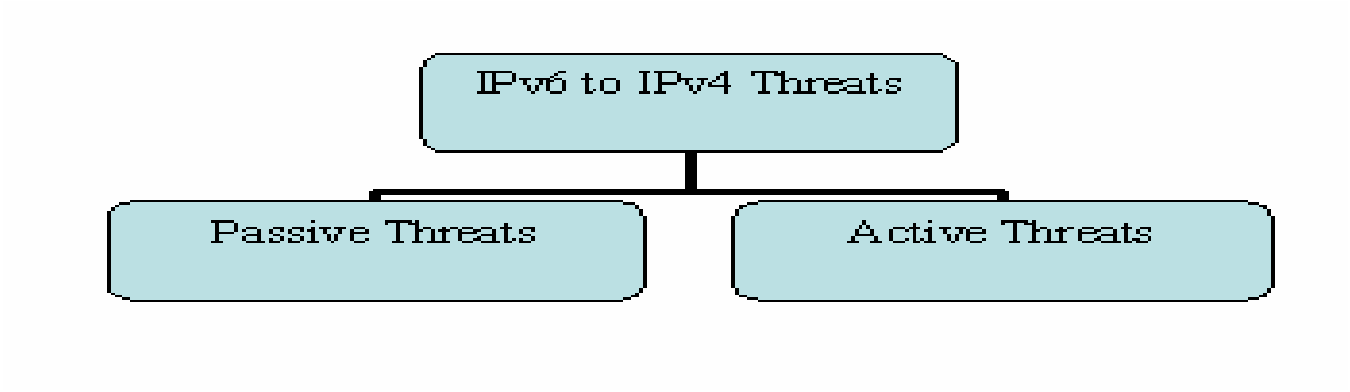

Fig.XII.The various types of IPv6 to IPv4 Threats.

Networks threats are broadly divided into two types.The first type of threat is passive threat and active threats.The term passive indicates that the attacker does not attempt to perform any modifications to the data [15].In fact,this is also why passive attacks are harder to detect, where the active attacks are based on modifications of the original message in some manner or in creation of a false message.

## IV. A brief overview of IPv6 Automatic Tunneling and Configuration tunneling mechanism considerations.

Several approaches to transition from IPv4 to IPv6 networks exist. These approaches are broadly divided into three following types[Ref-Figure-1].

1. Dual Stack, 2.Tunneling, 3.Translation.

### *A.IPv6 and IPv4 Threat Issues and Observations*

With regard to the IPv6 tunneling technologies and firewalls,if the Network designer does not consider IPv6 tunneling when defining security policy,unauthorized could possibly traverse the firewall in tunnels.This is similar to the issue with instant messaging(IM)and file sharing applications using TCP port 80 out of organizations with IPv4.According to some transition issues automatic tunneling mechanisms are susceptible to packet forgery and DOS attacks.

1.With regard to IPv6 Tunneling technologies and Firewalls,if the network designer does not consider IPv6 Tunneling when defining security policy,unauthorized traffic could possibly traverse the firewalls in tunnels.This is similar to the issue with instant messaging(IM) and file sharing applications using TCP port 80 out of organizations.

2.All ready we know that an automatic tunneling mechanisms are susceptible to packet forgery and DoS attacks.These risks are the same as in IPv4,but increase the number of paths of exploitation for adversaries.

3.According to the Network designer while deploying Automatic tunneling or Configuration tunneling,the tunneling overlays are considered non broadcast multi-access (NBMA) networks to IPv6 and require the network designer to consider this fact in the network security design.

4.An Automatic Tunneling with DoS threats and third parties has introduced by Relay translation technologies. These risks do not change from IPv4,but do provide new avenues for exploitation [13], either for external customers or internal customers the relays avenues can be limited by restricting the routing advertisements.

5.IPv6 to IPv4 and translation and relay techniques can defeat active defense trace back efforts hiding the origin of an attack.

6.Translation techniques outlined for IPv6 have been analyzed as shown to suffer from similar spoofing and DoS issues as IPv4 only translation technologies [14].

7.Static IPv6 in IPv4 Tunneling is preferred because explicit allows and disallows are in the policy on edge devices.

The below mentioned Figure-13 Shows a study that has been conducted by University of Mysore to estimate the IPv6 to IPv4 threats review with the help of automatic tunneling and Configuration tunneling issues.

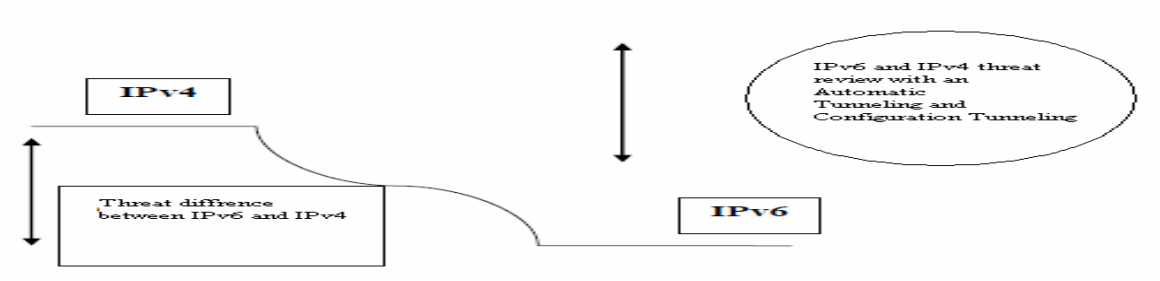

Fig. XIII.Threat differences between IPv4 and IPv6 with Tunneling Techniques.

## V.Prototype

### *5.1. Threat Analysis due to Transition Mechanisms*

Threat modeling (or analysis) is essential in order to help us to develop a security model than can focus or protecting against certain threats and manage the related assumptions.One methodology to discover and list all possible security attacks against a system is known as attack trees.To create an attack tree we represent attacks against a system in a tree structure, the attack goals as root nodes and the different sub goals necessary to achieve them as their leaf nodes.Figure-14 represents the general threat categories we have identified against network convergence architectures namely attack on the network processes are responsible for IPv6 transition,Dual stack,Automatic tunneling and Configuration tunneling threats.Dual Stack threats are totally different from the IPv6 Tunneling techniques like an automatic tunneling and Configuration tunneling,manually configured tunneling,Static tunneling etc.As we have discussed there are large number of transition mechanisms to deploy IPv6 but broadly be categorized into,Dual Stack,Tunneling(Automatic,Manual Configuration),and Translation Header.

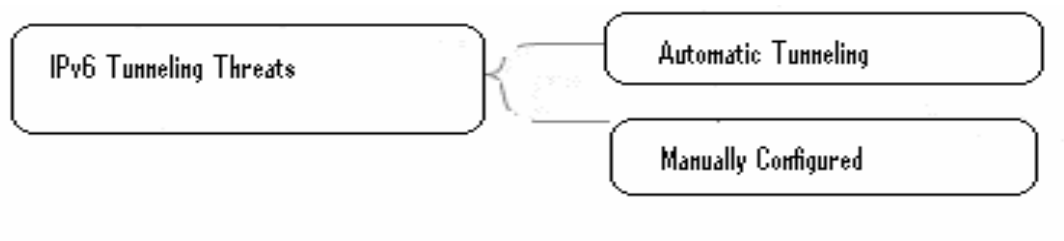

Fig.XIV.General Threat categories for IPv6 Tunneling.

The problems are identified when IPv6 is tunneled over IPv4 encapsulated in UDP as UDP is usually allowed to pass through NATS and Firewalls [59].Consequently allowing an attacker to punch holes with in the security infrastructure.The First and Second authors of this paper recommends that if the necessary security measures cannot be taken ,tunneled traffic should be used with caution if not completely blocked.To provide ingress and egress filtering of known IPv6 tunneled traffic, perimeter firewalls should block all inbound and outbound IPv4 protocol 41 traffic.For model we have identified that several attacks lead to other attacks which we have previously included and analyzed. These are represented in the tree as identical nodes in different locations.

*5.2. Security IPv6 Deployment for Automatic Tunneling and Configuration Tunneling.*

*A.Specific of IPv6 Tunneling Deployment*

Considering the issues surrounding IPv6 firewalls,Figure-15 demonstrates how all traffic originating from the Internet must be split up into its corresponding protocols.Each protocol must then be inspected and filtered independently based on a consistent policy before being forwarded to their respective destinations. Under most circumstances, the deployment model will be much more complex and will probably consist of some hybrid deployment structure, which may include some element of tunneling. For these situations, the principle should remain the same but the model should be adapted accordingly.

*D.Avoid IPv6 Tunneling or be aware of the security consequences.*

I.6 to 4 does not support source address filtering.,2.Teredo punches holes into the NAT device,3.Any Tunneling

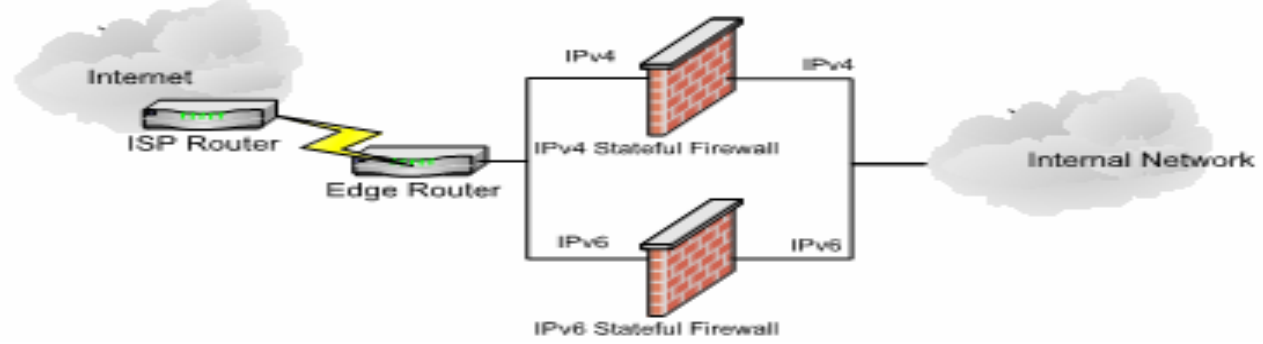

Fig.XV.Basic Security model for IPv6 Tunneling types.

*B. Predicted Security models for IPv6 Tunneling.*

Research is pushing the way towards reducing the restrictions that are preventing widespread deployment of Insect.Advances in the development of a PKI solution [49] to offer basic and advanced certification services,supplies a solution to allow client systems or end entities in one administrative domain to communicate securely with client systems or end users in another administrative domain.This can be extended to support multi-domain IPv6 scenarios through the deployment of cross-certification modules, thus reducing the key management problems. In addition to this, policy-based management systems are being implemented to solve the challenges presented by large-scale IPSec policy deployment across many network elements through the use of a centralized policy server which controls policy targets [50].These advances are in their preliminary stages but demonstrate the options available to start solving the circumstances where Protocol 41 is not blocked it can easily be detected and monitored by the open-source IPv4 IDS Snort.During the development of the IP6-to-IPv4 threat limitations of widespread IPSec deployment within IPv6.

C.*The more complicated the IPv6 transition/coexistence becomes the greater danger that security issues will be introduced either*

1.In the mechanisms themselves, in the interaction between mechanisms or by introducing unsecured paths through multiple mechanisms.

mechanism may be prone to spoofing, 4.With any tunneling mechanism we trust the relay-servers[60].

*E.Block IPv6 Tunneling Protocols.*

The networking and security communities have invested time and energy in ensuring that IPv6 is a security-enabled protocol. However, one of the greatest risks inherent in the migration is the use of tunneling protocols to support the transition to IPv6.These protocols allow the encapsulation of IPv6 traffic in an IPv4 data stream for routing through non-compliant devices.Therefore, it's possible that users on your network can begin running IPv6 using these tunneling protocols before you're ready to officially support it in production.If this is a concern,block IPv6 tunneling protocols (including SIT, ISATAP, 6to4 and others) at your perimeter.To block IPv6 Tunneling protocols we have to do two importantly configure the network by using upgrade our edge firewall, proxy and IDS to include IPv6 and tunneled IPv6 functionality, Drop all outbound IPv4 based UDP traffic with source or destination port 3544 and IPv4 protocol 41 packets[60].Along with these it includes leading threats of IPv4 DoS and DDoS attack.IPv6 DoS are related to Neighbor discovery(ND) protocol.ND includes various problems like address resolution, Neighbor unreachability detection, Duplicate address detection, and router discovery.These types of threats are controlled by IPSec[60].DDOS attacks are based on four representative modes like TCP-flood ,UDP-flood,ICMP-flood, Smurf attack.TCP involves three way handshake mechanism of the TCP protocol.The attacking node sends a series of SYN request to the victim with spoofed address.The victim will send SYN/ACK as response and wait some time for an ACK.Because of spoofed source address there is no ACK return. It causes the connection queue and memory buffer to fill up [62].

*F.Specific for Teredo: Network*

To restrict the outgoing traffic (white listing) then block UDP port 3544,then for Windows OS disable the teredo client, to disable the teredo client we have to use command like: netsh interface teredo set state disabled.To register teredo client the specific directory is

HKEY_LOCAL_MACHINE|SYSTEM\CurrentControlSet\Services\Tcpip6 \parameters [60].

*5.3. IPv6-Threats –IPv6 Tunneling.*

In Tunneling based methods,when a tunnel end point receives an encapsulated data packet,it decapsulates the in Figure-16,the target of attacks can be either a normal IPv6 node or the tunnel end point [25]

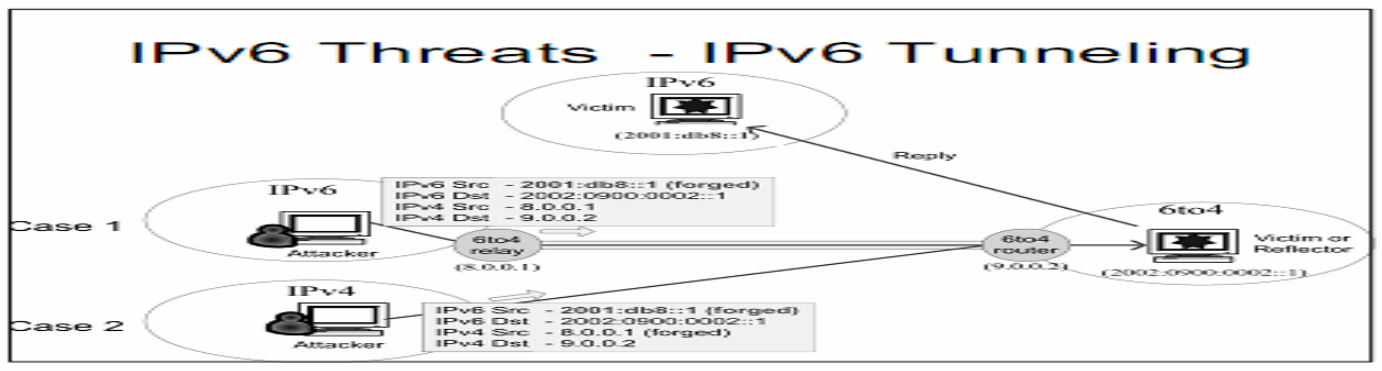

Fig.XVI.Security issues of various tunneling types.

*5.4. The Security issues in IPv6 tunneling are as follows.*

***1. The hard to trace back :***

Case-1:**IPv4 networking node can make an attack on IPv6 node(network):**The attackers(hackers) in IPv4 networks can make an attack on the IPv6 nodes through the 6to4 router(tunnel) end point by forwarding a spoofed encapsulated messages(Packets).Therefore here in this situation it is very difficult to trace back(Refer Fig-16).

Case-2:**IPv6 networking node can make an attack on IPv6 network (node):**In this type the hacker in IPv6 networks can make an attack on the IPv6 network through 6-to4 relay end point and 6-to4 router by sending a spoofed encapsulated packets.In this case also its very difficult to trace back.(Refer Fig-16)

2. **Potential reflect- DoS attack on Destination Host (Refer-Fig-16):**The hackers in the IPv4 networks can make a reflect–DoS attack to a normal IPv6 network (node) through the 6-to-4 router (tunnel) end point by sending the encapsulated packets with the spoofed IPv6 source address as the specific IPv6 node.

3. **Cheat by a Hacker with the IPv6 Neighbor Discovery (ND) message:** Whenever IPv4 network is treated as the link layer in tunneling technology,the hackers in the IPv4 networks can cheat and DoS attack the tunnel end point by sending encapsulated IPv6 neighbor discovery (ND)messages with a spoofed IPv6 link local address.The automatic tunneling techniques like 6-to 4 and Teredo get the information of remote tunnel end point from the certain IPv6 packets.

4. **Distributed Reflection DoS:** This type of attack can be performed if the very large number of nodes whenever involved in the sending spoofed traffic with same source IPv6 addresses.

packet and sends it to the other local forwarding scheme. The security threats in tunneling mechanisms,take IPv6 over IPv4 tunnel are mostly caused by the spoofed encapsulated packet sent by the attackers in an IPv4 networks.As shown

5. If the Destination host generates replies by using TCP SYN ACK ,TCP RST,ICMPv6 Echo reply,ICMPv6 Destination unreachable etc):In this case of attack the victim host is used as a reflector for attacking another victim connected to the network by using a spoofed source(Refer Fig-16).

6. **Spoofing in IPv4 with 6 to 4:** In this type of attack 6 to 4 tunneling spoofed traffic can be injected from IPv4 into IPv6.The IPv4 spoofed address acts like an IPv4 source, 6 to 4 relay any cast (192.88.99.1) acts like an IPv4 destination. The 2002::spoofed source address acts like an IPv4 destination.

7. IPv6 source address and the valid destination is IPv6 destination [28][Refer fig-17].

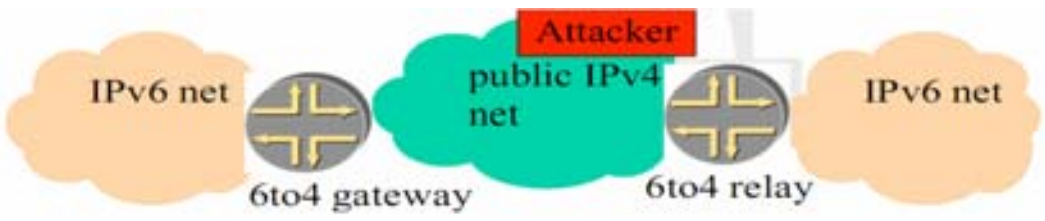

Fig.XVII.Spoofing in an IPv4 with 6 to 4.

7.**Theft of Service:** During the IPv6 transition period, many sites will use IPv6 tunnels over IPv4 infrastructure. Sometimes we will use static or automatic tunnels. The 6 to 4 relay administrators will often want to use some policy limit the use of the relay to specific 6 to 4 sites or specific IPv6 sites.However some users may be able to use the service regardless of these controls by configuring the address of the relay using its IPv4 address instead of 192.88.99.1 or using the router header to route the IPv6 packets to reach specific 6 to 4 relays.

8.**Attack with IPv4 broadcast address:** In the 6 to 4 mechanisms, some packets with the destination addresses spoofed and mapped to their broadcast addresses of the 6to4 or relay routers are sent to the target routers by the attackers in the IPv6 network.In this case also 6 to 4 or relay routers are attacked by the broadcast addresses.

The security issues in tunneling mechanisms can generally limited by investigating the validness of the source/destination address at each tunnel end point.Usually in tunneling techniques it is easier to avoid ingress filtering checks. Sometime it is possible to send packets having link-local addresses and hop-limit=255,which can be used to attack subnet hosts from the remote node, but it is very difficult to deal with attacks with legal IP addresses now[26].Since the tunnel end points of configuration tunnels are fixed, so IPSec can be used to avoid spoofed

attacks[29].IPv6 Security issues even though it has provided lot off features but its known that automatic tunneling are dangerous as other end points are unspecified because it's very difficult to prevent automatic tunneling mechanisms DoS/reflect-DoS attacks by the attackers in IPv4 network.

## VI. Recommendations

The conceptual ease of Tunneling mechanism deployment Configuration Tunneling.In addition to this,a general guideline is presented for a network administrator to take through each stage of deployment.

A.Immediate Actions to take before IPv6 deployment

1.When the Tunneled IPv6 is encapsulated in the following ways.

1.1:By using an IPv4 Header:Administrators who have not deployed IPv6 must first ensure that it is not being maliciously used without their knowledge.We know that 6to4,ISATAP,Tunnel Broker traffic is IPv6 traffic tunneled using an IPv4 header that has the IP protocol field set to 41.To protect from such traffic filter all the traffic with the IP protocol set to 41 set in an IPv4 header will prevent known IPv6 traffic from being tunneled within IPv4,thus preventing any back doors from being created within the network However,tunnels can also be set up over UDP,HTTP(port and so on,so the author recommends to use an IDS to carefully detect and monitor all tunneled traffic for instances of IPv6 traffic.

1.2:By using an IPv4 header and a UDP header:If is an IPv6 traffic is a IPv6 teredo traffic[Also called as(IPv4 network address translator(NAT-T) traversal (NAT-T) for IPv6 has resulted in it becoming one of the most popular transition methods. This Journal paper presents an overview of the protocols and technologies needed to secure current IPv6 Tunneling deployment, including basic security models, in addition to investigating and predicting future security models.The following section helps us to summarize all the key recommendations made throughout this paper to avoid IPv6 Tunneling techniques threats like Automatic and provides address and automatic tunneling for IPv6 connectivity across IPv4 Internet even when the IPv6/IPv4 hosts are located behind one or multiple IPv4 NAT'S]assignment and tunneled using an IPv6 header and a UDP port 3544.To protect from such Treed traffic drop(filter)all the traffic with the Source or Destination UDP port to set to 3544.The below figure-18 shows a figure of Teredo traffic(NAT-T)

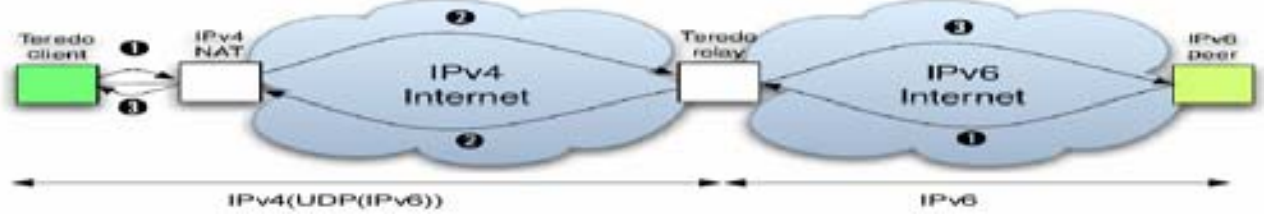

Fig.XVIII.Teredo Traffic (NAT-T Traffic).

**1.3:By using an 6 to 4 static tunneling** 6 to 4 is an address assignment automatic tunneling technology that is used to provide IPv6 connectivity between IPv6 sites and hosts across the IPv4 internet] instead of the specified tunneling techniques[Ref.fig-19.]

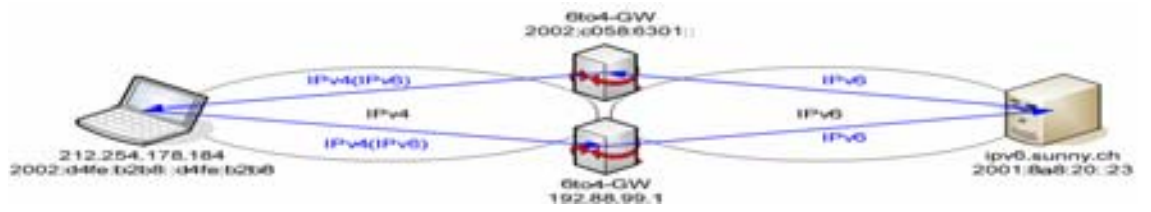

Fig.XIX.6-to-4 Static Tunneling.

## VII.The Current and future Innovative research challenges of IPv6 threat issues for Researchers.

This paper has not considered the overall threat review of IPv6 for all the aspects like dual stack,tunneling mechanisms,and Header Translation which are large and complex topics.To provide a complete overview of IPv6 security this paper should be in conjunction with the IPv6 to IPv4 threat review with tunneling considerations.The most important area to move forward with in IPv6 security is the extension of current IPv6 Firewalls and Network tools to test them(IPv6 packet constructors ,IDS's and so on).This will allow more users to adopt IPv6 without being paranoid about their openness to attack.

Before formulating analysis,we have proposed(formulated) several innovative research challenges.Presently there have been plenty of studies done on the research about basic security issues of IPv6, threat issues of IPv6, however there are still so many problems not yet resolved yet,calling for great challenges ahead.The innovative research challenges of IPv6 threat issues are as follows.

1.**Notion of the system identification within an organization:** With the advent of privacy extensions and the size of IPv6 ranges in use, identifying systems within an organization and in particular identifying mis behaving.

2.**Transition mechanisms from IPv4 to IPv6:**The current research on the basic transition mechanisms mostly focus on the situation of IPv6 over IPv4.Due to advanced deployment of IPv6,the IPv4 networks may also be separated by IPv6 ones. We are using only few kinds of method in this situation like IPv4 configuration tunnel and DSTM,more research on IPv4 over IPv6 transition methods is necessary.

3.Increased dependence on multicast addresses in IPv6 could have some interesting implications with flooding attacks. For example all routers and NTP servers have site specific multicast addresses. Can we use site specific multicast addresses to create an amplification attacks as similar as to the smurf attacks in IPv4.

4.We know that neighbor discovery is a new addition to the IPv6 to replace ARP and RARP of IPv4 and also it is an essential component of a well-run IPv6 network it should be tested from a security point like a neighbor-discovery cache fall victim to a resource starvation attack in any of the currently deployed neighbor discovery implementations. Can the CPU of a device be exhausted by processing information of IPv6 neighbor discovery?

5.IPv6 is new and security information on the protocol is not widespread, it is the opinion of all the authors that a large number of dual stack hosts may be more exposed to attack with IPv6 than in IPv4.

6.With a new IPv6 header configuration, new extension headers, and ICMP message types there may be a several novel ways to deal with flooding attacks.

7.**Scenario Analysis:**Typical Scenario analysis is still in progress. Some of them are in draft mode, such as enterprise network analysis along with this other possible scenarios should also be analyzed to support for next coming future wireless technologies.

8.**Support of Any cast,multihoming,multicast and Mobility:**All the research on basic transition mechanisms and analysis of typical transition scenarios normally focus on network connection. More effort should be made for the long process IPv6 transition to support multihoming, mobility, any cast and multicast.

9.**Security considerations**:All the IPv6 tunneling techniques are introducing more security however these problems cannot be settled or solved now a days. Besides the IPv6 firewall technology is also a good innovative topic for the future research.

10.**Difficult to identify Software and setup:**The various initialization of protocols of different transition issues like dual stack ,tunneling issues like automatic tunneling and configuration tunneling and header translation security make the chosen and setup of suitable IPv6 transition mechanisms difficult and more complex.A Standard way to discover and setup the software's for connecting the IPv6 networks across IPv4 only network and vice versa is needed for the interoperation of IPv4 and IPv6.

## VIII.Conclusion

This paper has shown both benefits and drawbacks from a security point of view.Many of the IPv4 threats (attacks)are similar to the IPv6 threats (attacks) but they are different in the way they are applied.All the tunneling techniques described here are useful in one way or another however they have different usage according to the type of network and the intended use of the tunnel.This paper outlines many of the common known threats against IPv4 and IPv6 and then it compares and contrasts how these threats issues or similar ones,might affect an IPv6 network.Automatic tunneling is also useful to provide hosts without support of their ISP with IPv6.Reading this paper should also stimulate further innovation ideas regarding the further research in IPv6 security.This paper has also shown IPv6 has both benefits and drawbacks from a security perspective. Many of the attacks applicable to principles of IPv6, but different in the way that they are applied. Sagacity of the presence of IPv6 and its corresponding transition methods is often enough to arm network administrators with enough information to thwart common attacks.This paper also has introduced us to the security issues and candidate best practices surrounding the introduction of IPv6 into a network with or without IPSec.Due to the prevalence of current Internet, the transition from IPv4 to IPv6 couldn't be accomplished in a short time.Besides the scarcity of IPv6 key applications IPv6 key applications makes no enough impetus to deploy IPv6 network.As a result, the transition to IPv6 is a long process.Threat estimation of IPv6 automatic and configuration tunneling can provide powerful security issues faster.As for UoM,Manasagangothri; the threat issues transition model is relatively very easy to understand tunneling threats.

**Acknowledgment**

The First author would like to thank Dr.Manjaiah.D.H, Reader,Mangalagangothri,Mangalore University, for his valuable guidance and helpful comments throughout writing of this journal paper.This research journal paper has been supported by Department of Studies in Computer Science,Manasagangothri,University of Mysore,and Department of P.G.Studies and Research in Computer Science,Mangalagangothri,and Mangalore University and also this Journal paper is also based upon works supported by Department of P.G.Studies and Research in Computer Science,Mangalagangothri,Mangalore University,under the University Grants Commission (UGC),New Delhi.

**References**

[1].Dr.Manjaiah.D.H. Hanumanthappa.J,2008, A Study on Comparison and Contrast between IPv4 and IPv6 Feature sets.In Proceedings of ICCNS'08, 2008,Pune,297-302.

[2].Dr.Manjaiah.D.H. Hanumanthappa.J. 2008,Transition of IPv4 Network Applications to IPv6 Applications, In Proceedings of ICETiC-09,2009, S.P.G.C.Nagar,VirudhaNagar-626 001,Tamil Nadu,INDIA-35-40.

[3].Dr.Manjaiah.D.H. Hanumanthappa.J. 2009,IPv6 over Bluetooth: Security Aspects, Issues and its Challenges, In Proceedings of NCWNT-09,2009, Nitte -574 110,Karnataka,INDIA –18-22.

[4].Dr.Manjaiah.D.H.Hanumanthappa.J. 2009,Economical and Technical costs for the Transition of IPv4–to- IPv6 Mechanisms [ETCTIPv4 to ETCTIPv6], In Proceedings of NCWNT-09, 2009,Nitte -574 110, Karnataka,INDA-12-17.

[5].S.Deering and R.Hinden,"Internet Protocol Version 6(IPv6) Specification", RFC 2460, December 1998.

[6].Silvia Hagen. 2002.IPv6 essential. New York: O'Reilly.

[7].Joseph Davies. 2003.Understanding IPv6.Washington: Microsoft Press.

[8].Dr.Manjaiah.D.H. Hanumanthappa.J.2009 An Overview of Study on Smooth Porting process scenario during IPv6 Transition (TIPv6),in Proceedings of IEEE IACC-09, 2009, Patiala, Punjab,INDIA-6-7,March-2009-2217-2222.

[9].S.Tanenbaum,"Computer Networks",Third Edition, Prentice Hall Inc., 1996, pp. 686,413-436,437-449.

[10].Behrouz A.Forouzan, Third Edition,"TCP/IP Protocol Suite"

[11].B.Carpenter and K. Moore,"Connection of IPv6 Domains via IPv4 Clouds,"RFC 3056, Feb.2001.

[12] P.Savola and C.Patel,"Security Considerations for 6to4," RFC 3964, Dec.2004.

[13].P.Savola,"Security Considerations for 6to4"(October 2003),at http://www.ietf.org/internetdrafts/draft-ietf-6ops-6to4-Security-00.txt.
[14].Okazaki,A Desai,"NAT-PT Security Considerations"(June 2003),at http://www.ietf.org/internetdrafts/draft-okazaki-v6ops-natpt-Security-00.txt
[15].Atul Kahate,"Cryptography and Network Security",Tata McGraw-Hill, 2003, pp-8-10.
[16].Kurose.J.& Ross .K.(2005) Computer Networking: A top-down approach featuring the Internet .3rd ed,(Addison Wesley).
[17].Deering.S.Hinden.R.(1998),Internet Protocol Version 6(IPv6) Specification .http: /// www.ietf.org/rfc/rfc2460.txt
[18].RFC 2553 –Basic Socket Interface Extensions for IPv6.
[19].Gilligan.& Nodmar .E.(1996) Transition Mechanisms for IPv6 Hosts and Routers.
[20].Silvia Hagen, 2002.IPv6 essential .New York: ORielly.
[21].Joseph Davies .2003.Understanding IPv6.Washington: Microsoft Press.
[22].Ioan R,Sherali.Z.2003.Evaluating IPv4 to IPv6 Transition mechanism.IEEE,West Lafayette, USA,v (1):1091–1098.
[31].Narten, T., Nordmark, E. and W. Simpson," Neighbor Discovery for IP Version 6(IPv6)",RFC 2461,December 1998.
[32].Thomson.S.and T.Narten,"IPv6 Stateless Address Auto configuration", RFC 2462, December 1998.
[33].Wellington, B.,"Secure Domain Name System(DNS) Dynamic Update",RFC 3007, November 2000.
[34].Mankin, A.,"Threat Models introduced by Mobile IPv6 and Requirements for Security in Mobile IPv6", Work in Progress.
[35].Kempf, J., Gentry,C. and A. Silverberg,"Securing IPv6 Neighbor Discovery Using Address Based Keys(ABKs)",Work in Progress,June 2002.
[36].Roe, M.,"Authentication of Mobile IPv6 Binding Updates and Acknowledgments",Work in Progress, March 2002.
[37].Arkko, J.,"Manual Configuration of Security Associations for IPv6 Neighbor Discovery",Work in Progress, March 2003.
[38].P.Nikander, J.Kempf, and E.Nordmark,"IPv6 Neighbor Discovery (ND) Trust Models and Threats", RFC3756,May 2004.
[39].J.Mohacsi, IPv6 Security:Threats and Solutions, http://www.6net.org/events/workshop-2005/mohacsi.pdf.
[40].E.Davies, S. Krishnan and P. Savola, "IPv6 Transition/Co-existence Security Considerations", draft-ietf-v6ops-security-overview- 06.txt (work in Progress),Oct 2006.
[41].Alvaro Vives and Jordi Palet, IPv6 Distributed Security:Problem Statement,Proceedings of the 2005 Symposium on Applications and the Internet Workshops (SAINT-W'05),IEEE, 2005.
[42].Kaeo, et. al., 2006, IPv6 Network Security Architecture 1.0, NAv6tf, www.nav6tf.org.
[43].Abidah Hj Mat Taib, IPv6 Transition: Why a new security mechanisms model is necessary.
[44].Savvas Chozas, "Implementation and Analysis of a Threat model for IPv6 Host Configuration",September 2006.
[45].Savola P., Security implications and considerations of internet protocol version 6(IPv6), October 2005.
[46].Baker, F.,"Requirements for IP Version 4 Routers", RFC 1812,June 1995.
[47].Savola, P,"Security of IPv6 Routing Header and Home Address Options", Work in Progress,March 2002.
[48].Ferguson.P. and D.Senie,"Network Ingress Filtering:Defeating Denial of Service Attacks which employ IP Source Address Spoofing",BCP 38, RFC 2827, May 2000.
[49].Carpenter, B. and K. Moore,"Connection of IPv6 Domains via IPv4 Clouds ", RFC 3056,February 2001.
[50].J.Hagino and K.Yamamoto, IPv6 –to IPv4 Transport relay translator, RFC 3142, June 2001.
[51].Jim Bound, IPv6 Transition JITC, FT Huachuca,AZ, and October 8-9, 2002.
[52].S.Kent and R.Atkinson,"Security Architecture for the Internet Protocol ", RFC 2401,Nov,1998.
[53].Savola P.(2003)Security considerations for 6to4 http://www.ietf.org/internetdrafts draft-ietf-v6ops-6to4-security-00.txt.
[23].S.Convey,D.Miller,IPv6 and IPv4 Threat Comparison and Best Practice Evaluation(V 1.0),Presentation at the 17th NANOG, May 24, 2004.
[24].J.Mohacsi, Security of IPv6 from firewalls point of view "presentation on TNC2004 Conference,June 2004.
[25].E.Davies, S.Krishnan and P.Savola,IPv6 Transition /Co existence Security considerations,draft–ietf-v6ops-natpt-toexprmntl-03,October 20, 2005.
[26].IPv6 Security address report,Ray hunt,Associate Professor,21-23 March,2005.
[27].P.Savola and C.Patel,Security considerations for 6 to 4,RFC 3964, December 2004.
[28].IPv6 Security,Malta 6 DISS workshop,4-6 April 2006.
[29].R.Graveman, M.parthasarthy, P.Savola, and H.Tschofeing using IPSec to secure IPv6 –in IPv4 tunnels draft –ietf—v6ops-ipsec-tunnels-01, August, 25,2005.
[30].Kent,S. and R.Atkinson,"IP Authentication Header",RFC 2402, November 1998.

[54].B.Schneir,"Attack Trees", Dr.Dobb's Journal, pp.21-19,1999.
[55].B.Carpenter and K.Moore,"Connection of IPv6 Domains via IPv4 Clouds", RFC3056, Feb 2001.
[56].D.Atkins, and R.Austein "Threat Analysis of the Domain Name System (DNS)" ,RFC 3833, 2004.
[57].A.Barbir, S.Murphy, and Y.Yang,"Generic Threats to Routing Protocols ",IETF Draft draft-ietf-rpsec-routing-threats-07,2004.
[58].P.Argyroudis and D.O.Mahony,"Secure Routing for Mobile Ad hoc Networks"IEEE Communications Surveys and Tutorials,vol.7,no.3, pp.2¨C21,2005.
[59].P.S.E.Davies,S.Krishnan,(2006,March)IPv6Transition/Co-existencesecurityconsiderations.IETF internal Draft [Online],Available ://www.cs.columbia.edu/~smb/papers/v6worms.pdf.
[60].Daniel Stirnimann,IPv6(IPv6 Transition and Tunneling Specific Issues),September 25,2008.
[61].Xinyu Yang, Ting ma, Yi Shi,"Typical DoS/DDoS threats under IPv6, Proceedings of International Multi-Conference on computing in the Global information Technology (ICGI'07) March 2007.Page(s):50-55.
[62].J.Postel,"Internet Protocol",DARPA Internet Program Protocol Specification, RFC 791,Sept .1981.
[63].Jelena Mirkovic and Peter Reiher,A Taxonomy of DDoS attack and DDoS defense mechanisms,ACMSIGCOMM Computer Communication review, PP.39-53, April 2004.
[64].J.Postel,NCP/TCP Transition Plan,RFC 801,November 1981.

## AUTHORS PROFILE

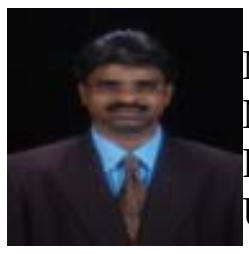

Mr.Hanumanthappa. is Lecturer at the DoS in CS,University of Mysore, Manasagangothri,Mysore-06 and currently pursuing Ph.D in Computer Science and Engineering,from Mangalore University under the supervision of **Dr.Manjaiah.D.H** on entitled "IPv6 Tunneling Issues for 4G Networks".His teaching and Research interests include Computer Networks, Wireless and Sensor Networks,Mobile Ad-Hoc Networks, Intrusion detection System, Network Security and Cryptography, Internet Protocols, Mobile and Client Server Computing, Traffic management, Quality of Service, RFID, Bluetooth, Unix internals, Linux internal, Kernel Programming, Object Oriented Analysis and Design etc.His most recent research focus is in the areas of Internet Protocols and their applications.He received his Bachelor of Engineering Degree in Computer Science and Engineering from University B.D.T College of Engineering,Davanagere,Karnataka(S),India(C),**Kuvempu University,Shimoga** in the year 1998 and Master of Technology in CS&Engineering from NITK Surathkal,Karnataka(S ),India (C) in the year 2003.He has been associated as a faculty of the Department of Studies in Computer Science since 2004.He has worked as lecturer at SIR.M.V.I.T,Y.D.I.T,S.V.I.T,of Bangalore.He has guided about 250 Project thesis for BE,B.Tech,M.Tech,MCA,MSc/MS.He has published about 15 technical articles in International, and National Peer reviewed conferences. He is a Life member of **CSI, ISTE,AMIE,IAENG**,Embedded networking group of **TIFAC–CORE in Network Engineering,ACM,Computer Science**

**Teachers Association(CSTA),ISOC,IANA,IETF,IAB,IRTG,etc**.He is also a **BOE Member** of all the Universities of Karnataka,INDIA.He has also visited **Republic of China** as a **Visiting Faculty of HUANG HUAI University of ZHUMADIAN,Central China,** to teach Computer Science Subjects like **OS and System Software and Software Engineering**, **Object Oriented Programming With C++,Multimedia Computing** for B.Tech Students.In the year 2008 and 2009 he has also visited **Thailand** and **Hong Kong** as a Tourist.

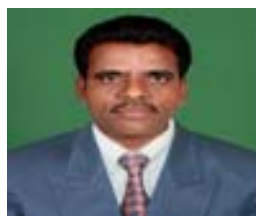

**Dr.Manjaiah.D.H** is currently Reader and **Chairman of BoS in both UG/PG in the Computer Science** at Dept.of Computer Science,Mangalore University,and Mangalore.He is also the **BoE Member** of all Universities of Karnataka and other reputed universities in India. He received **Ph.D degree** from University of Mangalore, **M.Tech. from NITK**, Surathkal and B.E.,from Mysore University. Dr.Manjaiah.D.H has an **extensive academic, Industry and Research experience**.He has worked at many technical bodies like **IAENG, WASET,ISOC,CSI,ISTE,and ACS**.He has authored more than-25 research papers in international conferences and reputed journals.He is the recipient of the several talks for his area of interest in many public occasions.He is an expert committee member of an **AICTE** and various technical bodies.He had written **Kannada text book**,with an entitled,**" COMPUTER PARICHAYA**",for the benefits of all teaching and Students Community of Karnataka.Dr.Manjaiah D.H's areas interest are **Computer Networking & Sensor Networks, Mobile Communication, Operations Research,E-commerce,Internet Technology and Web Prog ramming.**